 \documentclass{emulateapj}

\shorttitle{An expanding HI PDR flow in \object{GGD 14}}
\shortauthors{G\'omez et al.}

\begin{document}

\title{An expanding HI photodissociated region associated with the 
compact HII region G213.880-11.837 in the \object{GGD 14} complex}


\author{Y. G\'omez}
\affil{Centro de Radioastronom\'\i a y Astrof\'\i sica, UNAM,
    Apartado Postal 3-72, Morelia, Michoac\'an 58089, M\'exico.}
\email{y.gomez@crya.unam.mx}

\author{G. Garay}
\affil{Departamento de Astronom\'\i a, Universidad de Chile, Casilla 36-D, Santiago, Chile.}
\email{guido@das.uchile.cl}

\author{C. A. Rodr\'\i guez-Rico}
\affil{Departamento de Astronom\'\i a, Universidad de Guanajuato, Apartado Postal 144, Guanajuato, GTO 36240, M\'exico.}
\email{carlos@astro.ugto.mx}

\author{C. Neria, L. F. Rodr\'\i guez, V. Escalante and S. Lizano}
\affil{Centro de Radioastronom\'\i a y Astrof\'\i sica, UNAM,
    Apartado Postal 3-72, Morelia, Michoac\'an 58089, Mexico.}
\email{c.neria@crya.unam.mx, l.rodriguez@crya.unam.mx,
v.escalante@crya.unam.mx, s.lizano@crya.unam.mx}

\and
\author{M. Lebr\'on}
\affil{Departamento de Ciencias F\'\i sicas, Universidad de Puerto Rico, P.O. Box 23323, San Juan, Puerto Rico 00931-3323.}
\email{mayra.lebron3@upr.edu}

\begin{abstract}
We present high angular and spectral resolution HI 21~cm line observations
toward the cometary-shaped compact HII region G213.880-11.837 in the 
\object{GGD~14} complex. 
The kinematics and 
morphology of the photodissociated region, traced by the HI line emission,
reveal that the neutral gas is part of an expanding flow. 
The kinematics of the HI gas along the major axis of G213.880-11.837 shows that
the emission is very extended toward the SE direction, reaching
LSR radial
velocities in the tail of about 14 km~s$^{-1}$. The ambient 
LSR radial
velocity of the molecular gas is 11.5~km~s$^{-1}$, which suggests
a champagne flow of the HI gas. This is the second (after G111.61+0.37)
cometary HII/HI region known.

\end{abstract}


\section{Introduction}

The spatial morphology, abundance, and kinematics of the neutral gas around
compact HII (CHII) regions and its relation with other physical 
parameters remains poorly
studied for most HII regions. To date, only a few photodissociated 
regions (PDR)
have been observed at radio wavelengths with high angular resolution 
\citep[e.g.][]{LR97, GO98, GAR98, BR99, LRL01, Cap08}.
\object{GGD 14} is an active star forming region, part of the group of 
red nebulous objects GGD~12/13/14/15 \citep{GGD78} embedded 
in the Monoceros molecular cloud at a distance of d$\sim$1 kpc 
\citep{RMH80}.
There is an extended ($\sim$1.8 pc) bipolar outflow in the region, 
traced by CO emission with a major axis along the NW-SE direction 
\citep{Rod82, LHD90, Qin08}. This bipolar outflow seems to be excited
by a small region of free-free emission \citep[named VLA~7 by ][]{GRG00},
possibly a thermal jet \citep{eis00}. 
The powering source of this bipolar outflow is also associated with an 
H$_2$O maser \citep{Rod82}, a hydrogen molecular emission object 
\citep{FY04} and coincident in position 
with a class I/0 source, recently detected in the mid-infrared 
\citep[IRS~9Mc;][]{sat08}.

The \object{GGD~14} star forming region contains a cluster 
of radio sources dominated by the compact ($\sim$0.01 pc) cometary-shaped 
HII region (VLA~1; G213.880-11.837), which is excited by a B0.5 ZAMS star 
\citep{RMH80, KCW94, GO98, GRG00, GRG02} and is associated with 
IRAS~06084-0611.  A detailed kinematic study of the gas within the 
cometary HII region, presented by \citet{GO98}, shows that the ionized 
gas is undergoing a champagne flow with the head of the ionized
champagne flow located to the NW and the extended emission toward the SE. 
There is a velocity gradient along the major axis of the ionized
champagne flow with the LSR radial velocity increasing from 
$\sim$11 km~s$^{-1}$ at the head of the flow to $\sim$ 15 km~s$^{-1}$ 
at the tail. In a champagne flow the velocity at the head is expected to
be very similar to the velocity of the ambient molecular gas, and this is 
the case for the champagne flow in GGD~14 where the ambient velocity 
is about 11 km~s$^{-1}$ \citep{THR89, Qin08}. 

In addition of the ionized champagne flow, \citet{GO98} reported 
the existence of an unresolved PDR surrounding the ionized gas. 
The distribution of the
molecular gas in the \object{GGD~14} region
is complex with the presence of different velocity components
\citep{RMH80, GM86, THR89, LHD90, Ang96}. Recent 
$^{13}$CO(2{\bf $\rightarrow$}1) observations by \citet{Qin08}
toward the \object{GGD~14} region show that the molecular emission 
peaks at a velocity of $\sim$11.5 km~s$^{-1}$, that could be 
considered as the velocity of the ambient molecular gas.

In the present work we analyze new HI 21 cm line observations toward the
cometary HII region G213.880-11.837 in \object{GGD~14} with higher 
angular (15$^{\prime\prime}$) and 
spectral ($\sim$1~km~s$^{-1}$) resolution than the data presented by 
\citet{GO98}.
These new observations resolve the morphology 
of the HI region that surrounds the ionized material.
In \S 2  we present the observations, in \S 3 the results,
in \S 4 an interpretation of the HI observations, and in \S 5 the conclusions.

\section{Observations}

The HI 21~cm ($\nu$ = 1420.406 MHz) observations toward the star forming 
region \object{GGD~14} were made using the Very Large Array (VLA) of the 
NRAO\footnote{The National Radio Astronomy Observatory is operated by 
Associated Universities, Inc., under
cooperative agreement with the National Science Foundation.}.  
The observations were taken on December 23, 1998 in the C configuration
(under project AL459).
The total bandwidth of 0.781 MHz was centered at a 
V$_{LSR}$ of 12 km~s$^{-1}$, with 256 
channels 0.64 km s$^{-1}$ wide each (3.1 kHz). Hanning smoothing was applied
to the line resulting in a spectral resolution of $\sim$1.2 km~s$^{-1}$.
The flux density scale was determined 
from observations of the amplitude calibrator 0134+329, for which a flux density of 
15.87 Jy  was adopted. The phase calibrator was 0605-085 for which a 
bootstrapped 1.4 GHz flux density of 2.11~$\pm$~0.01 Jy was obtained. 
The data were edited and calibrated 
following standard procedures and images were made using the NRAO software AIPS.
Line images were made by subtracting the continuum
(line-free) channels from the visibility data using the task UVLSF.
The images were made with the task IMAGR, using the ROBUST parameter equal to zero \citep{BR95}, resulting in a synthesized beam of 
17$^{''} \times$ 14$^{''}$, P.A.=$-$26$^\circ$. 
The $rms$ noise level in a single spectral line channel, after
Hanning smoothing, is $\sim$2.0 mJy~beam$^{-1}$.

\section{Results}

The detected radio continuum emission at 21~cm, made with the line-free
channels, is marginally resolved 
(4${{\rlap.}^{\prime\prime}}$3$\times$2${{\rlap.}^{\prime\prime}}$9, 
P.A.=148$^\circ$), 
with a flux density of 72$\pm$2 mJy and 
peak position at $\alpha$(2000)= 06$^h$ 10$^m$ 50${{\rlap.}^s}$607 
$\pm$0${\rlap.}^s$004,
$\delta$(2000)=$-$06$^\circ$ 11$^\prime$ 50${\rlap.}^{\prime\prime}$10 
$\pm$0${\rlap.}^{\prime\prime}$05.
The HI 21~cm line, is detected both in absorption and emission in the 
velocity range from $\sim$1 to 20 km~s$^{-1}$, in agreement with 
previous results reported by \citet{GO98}. 

Figure~1 shows the HI 21~cm spectrum integrated over the whole source
in a box with size of $\sim$1${\rlap.}^{\prime}$3$\times$1${\rlap.}^{\prime}$3.
As was noted by \citet{GO98}, the HI absorption component is
considerably broad towards blueshifted velocities, showing three minima and
extending from velocities of $-$1 km~s$^{-1}$ to $\sim +$12 km~s$^{-1}$. 
In Table~1 we present the results of a Gaussian fit for the 
three HI 
absorption components plus a single Gaussian for the emission component 
(the fit is shown as a dashed-line in Figure~1).
In this Table we give 
the peak flux (S$_l$), the center velocity (V$_{LSR}$) and the FWHM 
(full width at half maximum) of the line ($\Delta$V) for each 
Gaussian component. The HI absorption 
components are centered at 
V$_{LSR}$ of $\sim$ 1, 6 and 11 km~s$^{-1}$ (see Table~1).
On the other hand, the HI emission is present in a velocity range from
$\sim$ 13 to 20 km~s$^{-1}$, with the maximum centered at
V$_{LSR}$=15.5 km~s$^{-1}$ (Figure~1 and Table~1).

Figure~2 shows individual channel images of the HI 21-cm line toward 
GGD~14.
The unresolved HI absorption appears coincident with the 1.4~GHz
continuum peak position \citep[named VLA~1 by][]{GRG00}, marked with a cross.
The triangle, towards the NE of the CHII region, indicates
the radio continuum peak position of the radio source VLA~7, which
has been proposed as the exciting source of the bipolar outflow
\citep{GRG00, GRG02} and is
coincident with the H$_2$O maser emission in the region \citep{RMH80}.
As observed in the HI line in emission, the PDR in GGD~14 appears
spatially resolved with a deconvolved
angular size of 45$^{\prime\prime} \times$40$^{\prime\prime}$, P.A.=34$^\circ$
($\sim$0.2 pc) and with a peak intensity spatially shifted
toward the SE side of
the cometary HII region 
(the PDR peaks at $\alpha$(2000)= 06$^h$ 10$^m$ 51${{\rlap.}^s}$3,
$\delta$(2000)=$-$06$^\circ$ 11$^\prime$ 52$^{\prime\prime}$
$\pm$ 2$^{\prime\prime}$, about 0.1~pc in projected separation from the 
peak of the CHII region).

\section{Discussion}

\subsection{Physical parameters of the HI line}

The individual HI 21~cm channels, toward GGD~14 (Figure~2), show an extended 
emission component toward the SE and a strong line absorption 
coincident with the cometary HII region. The PDR, traced by the HI region,
is thus being produced mainly by the B0.5 ZAMS star that also ionizes the 
cometary HII region (VLA1; G213.880-11.837).

It is possible to estimate the HI optical depth toward the CHII region 
using the HI absorption profile.
Assuming that the continuum source is located behind a homogeneous 
HI cloud, the observed line brightness temperature is given by
\citep[see][]{RW03}:
$$T_L(v)={\big(f_{HI}~T_{ex} - f_0~f_c~T_c\big)(1 - e^{-\tau_L(v)})},$$
where T$_{ex}$ is the excitation temperature of the HI, T$_c$ is the 
continuum brightness temperature of the CHII region at the frequency of the
HI line, f$_{HI}$ and f$_c$ are the beam filling factors of the HI and the 
continuum, f$_0$ is the fraction of the continuum source covered by the
HI cloud and $\tau_L(v)$ is the optical depth of the HI gas at a 
velocity $v$.
In this case, we consider that the continuum source is completely covered 
by the HI cloud, f$_0$=1. The estimation of the filling factor 
commonly uses a comparison between
the beam solid angle ($\Omega_B$) with the source solid angle ($\Omega_c$),
such that $f_c\simeq\Omega_c/\Omega_B$, but other geometric
factors can be included.  

Since the HI cloud shows a velocity gradient across the region (see Figure~2), 
we will assume that at the velocity of the absorption feature 
(V$_{LSR} \simeq$11 km~s$^{-1}$), we only have to take into account 
the effect of the absorbing gas, neglecting the contribution to the
emission (f$_{HI} T_{ex}\sim$0), since this occurs mostly at
other velocities. Then, the transfer equation at the
center of the absorbing line is,
$$T_L(0)={\big(- f_c~T_c\big)(1 - e^{-\tau_L(0)})}.$$
The resulting optical depth value is obtained using the expression:
$$\tau_L(0) = -\ln\Big(1 - \frac{|T_L(0)|}{f_c~T_c}\Big).$$
The beam-averaged brightness temperature of the continuum can be computed 
from the total flux density (S$_c\simeq$72 mJy), and the deconvolved 
angular size
measured in the 1.4 GHz continuum image 
($\theta_c \simeq 3\rlap.{''}5$), by the expression 
T$_c=688(S_\nu/\Omega_c)$, with $\Omega_c=1.13\theta_c^2$,
obtaining T$_c$=3,600 K. The filling factor is 
f$_c \simeq \theta_c^2$/$\theta_B^2$, where $\theta_B^2$ is the beam size 
($\sim 15\rlap.{''}4$), thus f$_c$=0.05. 
The observed line brightness temperature, at the center of the absorbing line
T$_L(0)$, for S$_L(0)$=$-$69~mJy and $\theta_B$=15$\rlap.{''}4$, is $-$177~K. 
Then the mean optical depth has a very large value $\tau_L(0)\simeq$5. 
As we note, the derived values for T$_L(0)$ and f$_c$T$_c$ are similar,
such that the ratio T$_L(0)$/f$_c$T$_c$ will be very close to unity,
making the estimation of the optical depth very uncertain. Moreover,
If we include the uncertainty in the estimated angular size and 
temperatures, the error in the mean optical depth will be around 60$\%$,
$\tau_L(0)\simeq$5$\pm$3.

A lower limit to the HI mean density can be set by assuming an optical
depth, ${\tau_L(0)}>>1$. This assumption is supported by our previous
estimate, $\tau_L(0) \sim 5$, but with large uncertainties. 
Then taking simply $\tau_L(0) \geq$1, T$_{ex}\simeq$300~K and 
the FWHM of the line $\Delta$V=2~ km~s$^{-1}$ (see Table~1) we can 
estimate a lower limit for the HI column density, computed in the standard 
way \citep{RW03},
$$\Biggl[\frac{N_{HI}}{cm^{-2}}\Biggl] = 1.8224 \times 10^{18}\Biggl[\frac{T_{ex}}{K} \Biggl] \tau_L(0) \Biggl[\frac{\Delta V}{km~s^{-1}}\Biggl],$$  
giving
N$_{HI} \geq$1.1$\times$10$^{21}$~cm$^{-2}$. This column density is 
consistent with that expected for PDRs \citep{ht97}.
We will further assume that the physical depth of the region
producing the HI absorption is approximately one half of the 
diameter of the HI in emission ($\sim$42$^{\prime\prime}$;
$\simeq$~0.2~pc). Therefore an estimate of the average neutral 
hydrogen density, $<n_{HI}> \simeq$ N$_{HI}$/L, where 
L$\simeq$0.1~pc, is $<n_{HI}> ~\geq$3.5$\times$10$^3$~cm$^{-3}$.
 
The neutral hydrogen mass associated to the HI emission
can be obtained assuming a spherical cloud with an HI radius of
21$^{''}$ (0.1~pc) and average hydrogen density 
$<n_{HI}> ~\geq$3.5$\times$10$^3$~cm$^{-3}$, from the expression  
$M(HI)= \frac{4\pi}{3} {R^3 ~<n_{HI}> m_{HI}}$, where $R$ is the radius
of the HI emission region, and 
$m_{HI}$ is the hydrogen atomic mass. We obtain that the HI mass 
M(HI)$\geq$0.4~M$_\odot$. This value 
should be taken as a lower limit, since the average hydrogen density
is also a lower limit. 
\citet{GO98} used a line emission model for the PDR to estimate the total 
HI mass, obtaining a value as high as 5~M$_\odot$. 
Molecular mass estimates, derived from CO, indicate that GGD~14 
has a molecular mass in the range from 640 to 1500 M$_\odot$ 
\citep{rid03, hig09}, which means that the PDR mass is very 
low ($<$1$\%$ of the total gas mass). Of course, the PDR around 
G213.880-11.837 extends only over $\sim$1$^{'}$, while the CO is detected 
over $\sim$5$^{'}$ \citep{hig09}. 

\medskip
\subsection{Kinematics of the HI gas in GGD 14}

The high optical depth toward the CHII region (${\tau_L(0)}\geq$1) supports
the existence of high density atomic gas,
where the cometary CHII region (G213.880-11.837) is embedded. 
The spatial extension of the cometary HI 21-cm emission
($\sim$42$^{''}$) is $\sim$10 times larger than the ionized
CHII region ($\leq$3.5$^{''}$ at 1.4~GHz).
In the champagne model a strong density gradient is required, with
the head of the cometary CHII region embedded in the densest part of the
molecular clump and the ionized gas expanding asymmetrically out of the
dense clump \citep{GL99}. For G213.880-11.837, \citet{GO98} 
interpreted 
the kinematics of the ionized gas, derived from the H92$\alpha$ line, as a
champagne flow due to the presence of a velocity gradient measured from
the head (located to the NW with V$_{LSR}\sim$11 km~s$^{-1}$) to the tail 
(located to the SE with V$_{LSR}\sim$15 km~s$^{-1}$).

Recent numerical results, show that an expanding H~II region should be 
an efficient trigger for star formation in molecular clouds if the mass 
of the ambient molecular material is large enough \citep{HI06}.
In the particular case of the GGD~14 complex, a dense molecular region 
should exist in the vicinity of G213.880-11.837, where recently a
cluster of low-mass pre-main-sequence stars has been found.
In addition to the radio continuum compact sources \citep{GRG00, GRG02}, 
there are infrared studies in the 2-100 $\mu$m range  
\citep{har85, hod94, FY04} that provide evidence for a 
recently formed cluster of sources in the vicinity of G213.880-11.837.
\citet{FY04} estimated the age of this cluster in the order 
of 2$-$5 $\times$ 10$^{6}$ yrs. 

The molecular gas detected toward GGD~14 shows a complex velocity
structure with the presence of a CO molecular outflow with the
major axis along the NW-SE
direction (P.A.=$-$60$^\circ$), and centered near the radio source VLA~7. 
In particular, we note that, in the CO(2{\bf $\rightarrow$}1) images 
\citep{LHD90, Qin08}, 
there is strong self-absorption of this line at V$_{LSR} \sim$11 km~s$^{-1}$, 
which correspond to the ambient molecular 
gas velocity. \citet{Qin08} shows a $^{13}$CO(2{\bf $\rightarrow$}1)
spectrum toward this region (not affected by self-absorption), that  
peaks at a velocity of $\sim$11.5 km~s$^{-1}$. In what follows we will
assume this value for the ambient molecular gas toward GGD~14. 
There is also NH$_3$ \citep{torr83, GM86, Ang96}, 
CS \citep{Ang96} and HCO$^+$ emission \citep{hea88}, toward GGD~14, 
with velocities in the range from 9 to 12 km~s$^{-1}$. These results 
point out that there is a dense region in the vicinity of the CHII region 
that could be responsible of the champagne morphology 
observed in both the CHII and the HI region. 

\medskip
\subsubsection{Champagne HI flow around G213.880-11.837?}

Figure~3 shows a near-IR continuum image at 4.5 $\mu$m of 
Spitzer Space Telescope\footnote{The Spitzer Space Telescope, is 
operated by the Jet Propulsion Laboratory, California Institute of 
Technology under a contract with NASA.} toward GGD~14. 
We used the processed Spitzer IRAC \citep{faz04} Basic Calibrated 
Data (PID: 6. PI: Giovanni Fazio),
which are available in the archive of the Spitzer Science Center. 
In this image the stellar cluster is clearly appreciated. We
overlay in this image the with integrated HI 21~cm line, from 1 
to 20 km~s$^{-1}$.  The stellar cluster
is mainly located toward the HI absorption region. We notice that 
the compact radio sources VLA~1 (G213.880-11.837) and VLA~7 have strong 
4.5 $\mu$m counterparts not observed at 2.2 $\mu$m  \citep{GRG02}, 
which suggests that these sources are deeply embedded. The
4.5 $\mu$m counterpart toward VLA~1 is first reported in this work, and 
that toward VLA~7 was recently detected by \citet{sat08}. 
The near-IR source IRS~9Mc reported by \citet{sat08} 
is localized near VLA~7, and shows a bipolar infrared nebula aligned in the
same direction as the bipolar molecular outflow \citep[NW-SE;][]{Rod82}. 
We also notice that several of the compact radio sources reported by 
\citet{GRG02}, indicated with crosses, have 4.5 $\mu$m counterparts.

Figure~3 shows the velocity integrated HI 21~cm line toward G213.880-11.837, 
which exhibits a cometary like morphology,
resembling (at much larger scales) that of the ionized gas at
high frequencies \citep{tof95, KCW94, GO98}.
The limb brightened PDR expected at the head of the cometary 
region G213.880-11.837 is not observed. This can be due to the 
HI absorption feature that complicates
the full imaging of the PDR or to the low angular resolution of the HI
observations. In the latter case, one can set a lower limit to the HI
volume density at the head of the cometary. Assuming a PDR size equal
to the beam size, $l = 1.5 \times 10^4 {\rm AU} (\theta_B/15^{''})( d/ 1kpc)$,
and a column density of the PDR ${\rm N}_{HI}\sim3\times10^{21}~{\rm cm}^{-2}$ 
such that the dust opacity is of order unity, 
$n_{HI} \ge {\rm N}_{HI}/l = 1.3 \times 10^4~{\rm cm}^{-3}$, 
which is a reasonable density for molecular clouds.  
Molecular line observations with high spatial
resolution are required to try to detect this high density
molecular gas at the head of the cometary.

In order to assess whether or not the HI emission is produced by a 
champagne flow, we made position-velocity diagrams of the HI emission along 
two perpendicular directions, centered on the CHII region 
(peak continuum emission).
The cuts are at position angles of +25$^{\circ}$ and $-$65$^{\circ}$ 
which correspond to the orientations of the minor and major axis 
of the ionized cometary CHII region \citep{GO98}.
Figure~4 shows the PV diagrams of the HI gas along these two 
directions. They both exhibit absorption and emission features.
The spatially unresolved HI line in absorption seen at the position of 
the CHII region is strikingly broad in velocity, ranging from 5 to 
13 km~s$^{-1}$.  We also see that at this position the HI emission reaches 
its maximum redshifted velocities with a value of about 18 km~s$^{-1}$. 
\citet{Qin08} have shown that the velocity of the ambient molecular 
gas toward GGD~14 is $\sim$11.5~km~s$^{-1}$. 
We interpret these two features as indicating the presence of an expanding 
shell of HI gas around the CHII region, with an
expansion velocity of about 6.5 km~s$^{-1}$.

The PV diagram along the position of $-$65$^{\circ}$ shows that 
the emission is very extended toward the SE direction, reaching 
velocities in the tail of about 14~km~s$^{-1}$. 
Since the ambient velocity of the molecular gas is 11.5~km~s$^{-1}$,
we interpret the PV structure seen SE of the CHII region as 
indicating a champagne flow of the HI gas. 
In a champagne flow the gas accelerates to 2 $-$ 3 times the sound speed
\citep[see Table~2 of][]{shu02}.
Since the observed velocity shift in
the ionized gas is $\Delta v \sim$ 15~km~s$^{-1} -$ 11 km~s$^{-1}$ = 
4 km~s$^{-1}$ \citep{GO98} while a shift of $\Delta v \sim$ 
20 $-$ 30~ km~s$^{-1}$ is expected, this ionized flow can be understood as 
a champagne flow only if the source is
inclined close to the plane of the sky, with an angle $ i < 11^\circ$,
and has a small opening angle. \citet{GO98} correctly interpreted
the kinematics of the ionized gas as a champagne flow but they did not
estimate the inclination angle of the flow with respect to the plane
of the sky, they only suggest that the tail should be going away to explain 
the redshifted velocities. In the present interpretation we propose that 
in fact the orientation of the ionized champagne flow should be close to 
the plane of the sky with the tail having the redshifted velocities. 
On the other hand, in a PDR, one
expects gas temperatures of $T \sim 500$~K, i.e., sound speeds 
$\sim$2~km~s$^{-1}$. Thus, for a champagne flow in the HI neutral gas, 
the expected velocity shift is of the order of $\sim$4 $-$ 6 km~s$^{-1}$.  
Thus, the observed velocity shift, $\Delta v_{HI} \sim$ 3 km~s$^{-1}$, is 
consistent with a champagne flow of the HI gas, with a large opening 
angle such that the observed velocity in the line-of-sight is small.  
Note that the scale of this HI flow is $\sim$10 times larger 
than the ionized champagne flow, therefore, the collimation of this 
HI neutral gas is probably much lower supporting the presence of
a larger opening angle.

In contrast, in the cut along the perpendicular
direction to the symmetry axis passing through the HII region
(P.A.=+25$^{\circ}$), the HI gas in emission is seen at both sides of
the cometary head and no velocity gradient is appreciated.
High angular resolution molecular observations are needed to
confirm this kinematics.
Another HII region
that shows a cometary morphology in both the continuum and HI
emission is G111.61+0.37, where a champagne interpretation was
also proposed \citep{LRL01}.

Hydrodynamical and radiative transfer models by \citet{Hos07} show 
that an inhomogeneous medium can change the time evolution of
expanding HII regions and PDRs (as compared with the evolution in a
homogeneous medium). In a medium with a radial density 
profile of the form $n(r)\propto r^{-w}$ with $0\le w<1$, the 
dense shell formed by 
the shock front shields the far-ultraviolet (FUV) radiation and 
prevents the expansion of the dissociation front beyond the shell. 
The CHII region is 
surrounded by a cold molecular envelope in this case. 
In a stepper density gradient with $w>1$, as expected in a ``champagne 
flow,'' shielding of the FUV radiation becomes inefficient and an 
extended PDR develops around the HII region. However, only 
if $w>1.5$, the H$_2$ dissociation front propagates through the 
ambient cloud. Since in G213.880-11.837 and G111.61+0.37 we observe extended  
PDRs around the HII regions, it is likely that steep density gradients
exist in both objects.
 
\medskip
\section{Conclusions}

We present high angular resolution (15$^{''}$) VLA observations
of the HI 21~cm line toward the compact HII region G213.880-11.837. The 
HI gas, produced by a photodissociated region around the CHII region, 
exhibits a cometary morphology with an
extension, along the major axis, of $\sim$45$^{''}$.  
The kinematics and morphology of the HI gas around G213.880-11.837 
indicates that
the neutral hydrogen gas is in expansion in a  
similar way as the ionized gas. 
In both cases the gas accelerates 
toward lower density regions and redshifted velocities (from 
$\sim$11 to 14 km~s$^{-1}$). The peak position of the HI absorption
coincides with the peak continuum emission of the CHII region (VLA~1; G213.880-11.837). 
This is the second (after G111.61+0.37)
cometary HII/HI region known. These observations suggest the existence
of champagne flows in PDRs associated with cometary HII regions.

\acknowledgments

We thank an anonymous referee for the comments and suggestions that
helped to improve this paper.
Y.G., L.F.R. and S.L. acknowledge support from DGAPA-UNAM and CONACyT Mexico.
G.G. acknowledges support from CONICYT projects FONDAP 15010003 
and BASAL PFB-06.
 




\clearpage



%
\begin{figure}
\centering
\includegraphics[scale=0.9, angle=0]{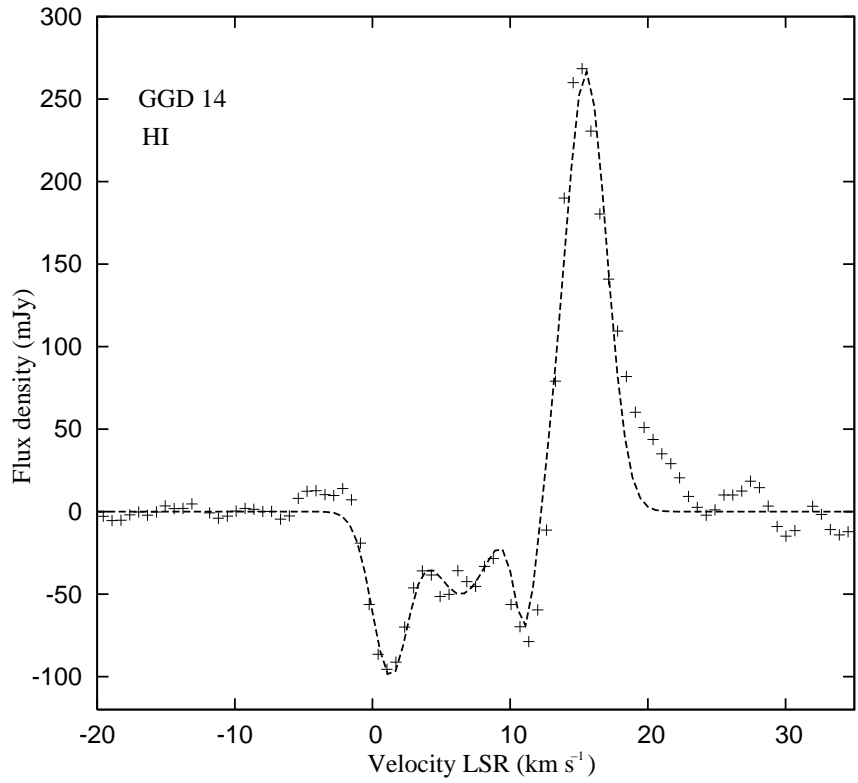}
\caption{Integrated spectrum of the HI 21~cm line emission
toward GGD~14. The dashed line shows the fit to the data
with four Gaussians (see Table~1).
}
\label{fig1}
\end{figure}

\begin{figure}
\centering
\includegraphics[angle=0,scale=0.9]{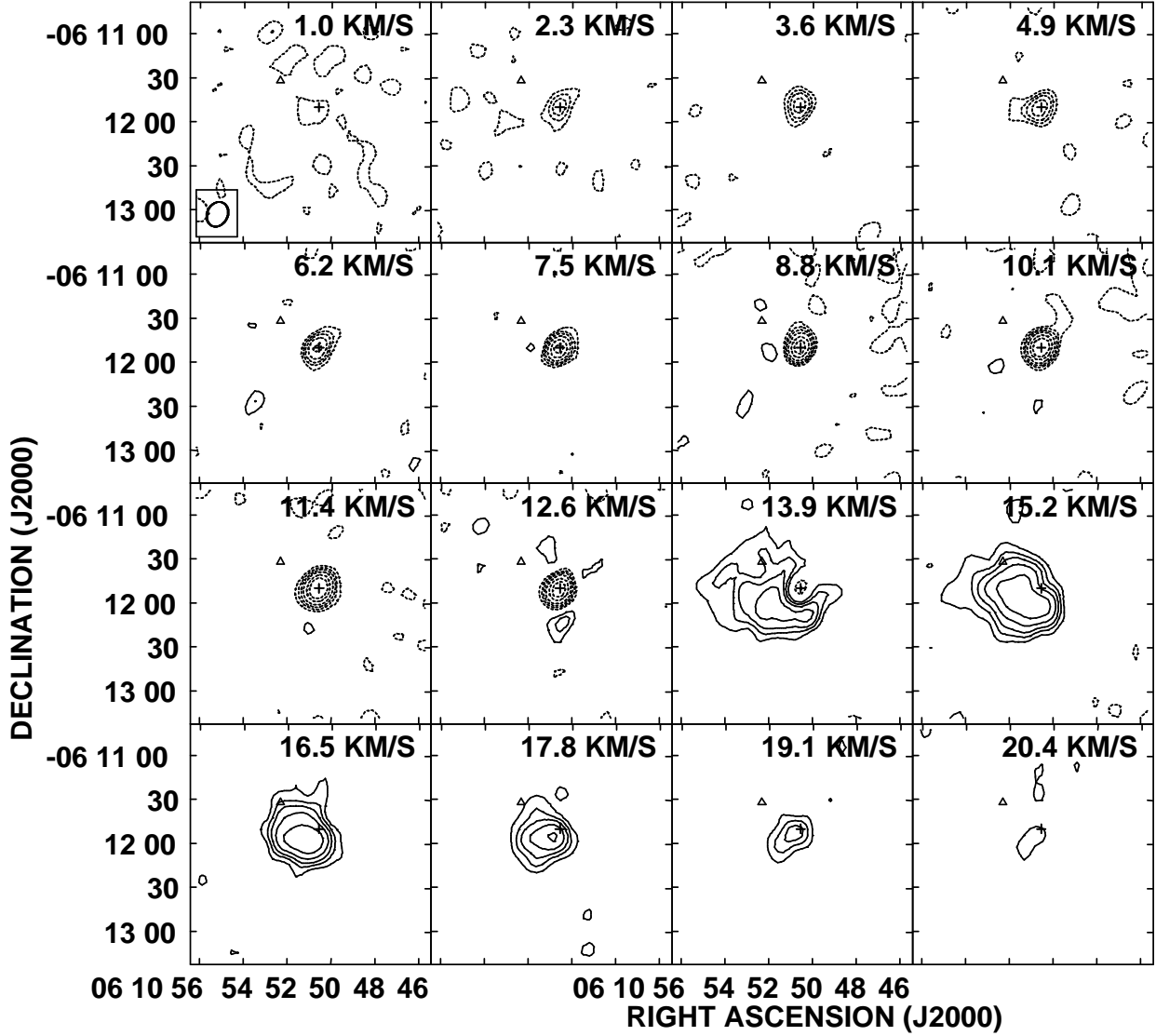}
\caption{Individual channel images of the 21~cm HI emission and absorption
toward the GGD~14 region. One out of every two channels
was plotted. Contour levels are -20, -15,
-10, -7, -3, 3, 5, 7, 10, and 15 $\times$ 2 mJy~beam$^{-1}$, the $rms$
noise of the images.  The continuous and dashed contours indicate emission
and absorption, respectively.
The cross and the triangle mark the position of the radio
continuum peak at 1.4 GHz of VLA~1 (G213.880-11.837) and VLA~7 \citep{GRG00}, 
respectively.
}
\label{fig2}
\end{figure}

\clearpage

\begin{figure}
\centering
\vspace{3cm}
\includegraphics[angle=0,scale=0.7]{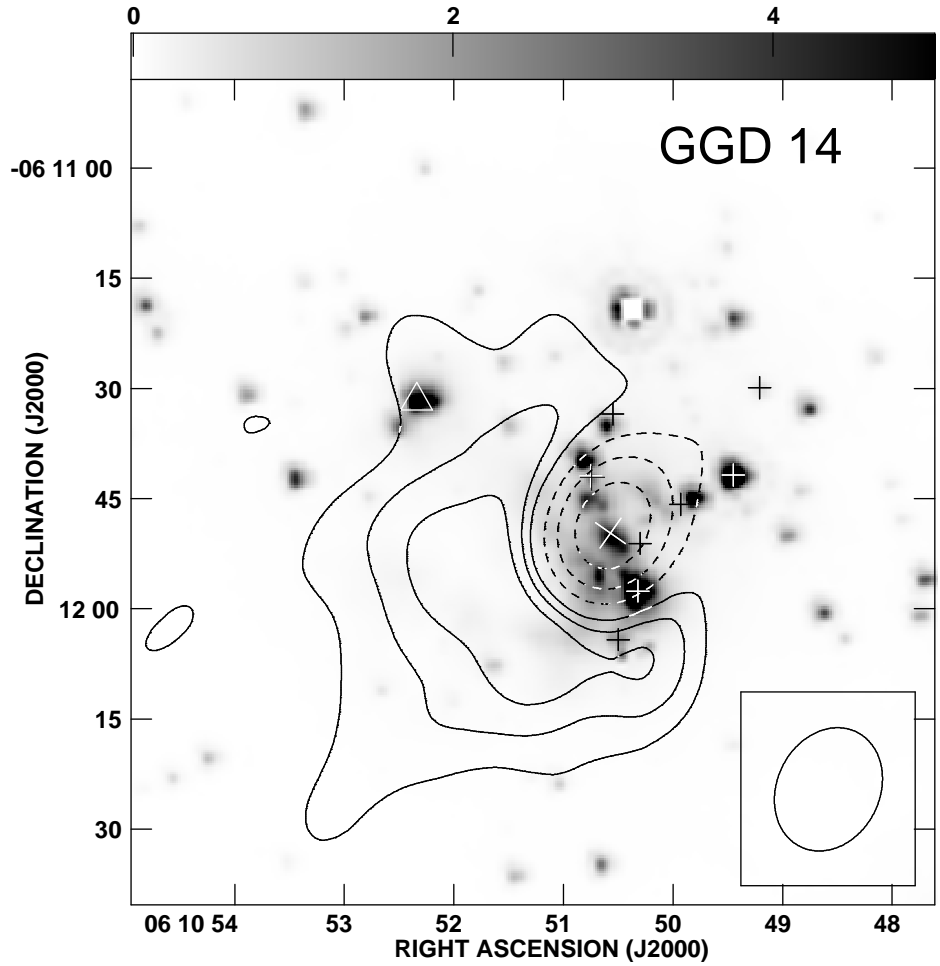}
\caption{Superposition of the near-IR Spitzer image at 4.5 $\mu$m 
(gray scale) on the VLA integrated HI line (from 2 to 20 km~s$^{-1}$) 
with solid (emission) and dashed (absorption) contours (-20, -10, -3, 3, 6
and 9 $\times$ 0.7 mJy~beam$^{-1}$~km~s$^{-1}$). 
The large $\times$ symbol indicates the radio continuum peak position 
of the HII region (VLA~1) and the triangle the peak position of VLA~7. 
The crosses mark the peak position
of the compact radio sources reported by \citet{GRG02}. 
}
\label{fig3}
\end{figure}

\clearpage

\begin{figure}
\centering
\vspace{3cm}
\includegraphics[angle=0,scale=0.5]{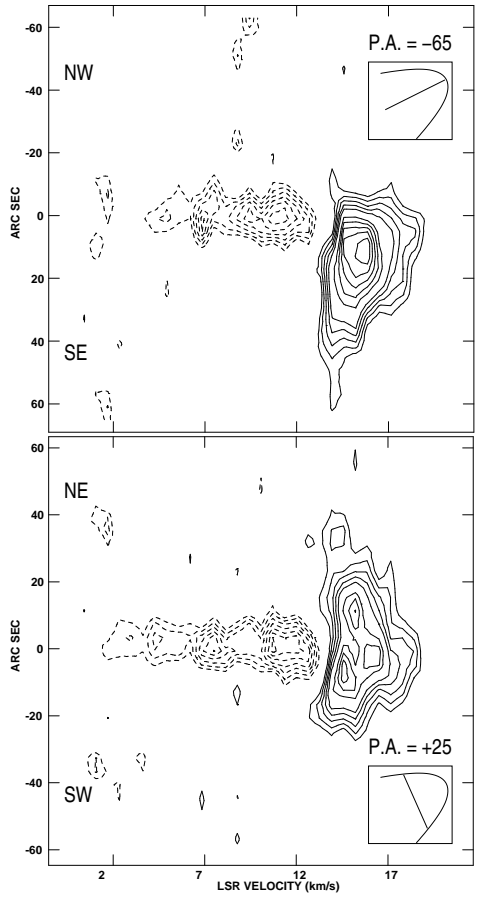}
\caption{Position velocity diagrams of the HI line toward GGD~14. Top:
cut along the symmetry axis of the cometary HI structure (P.A.=$-$65$^{\circ}$).
Bottom: cut along a perpendicular direction to the symmetry axis passing 
through the HII region (P.A.=+25$^{\circ}$). Contour levels are -24, -21, 
-18, -15, -12,-9, -6, -3, 3, 6, 9, 12, 15, and 18 $\times$ 2.5 mJy~beam$^{-1}$,
 the $rms$ noise in these images. The insets show the direction of the cuts
with respect to the cometary morphology.
}
\label{fig4}
\end{figure}

\clearpage
\begin{deluxetable}{rcc}
\tabletypesize{\tiny}
\tablecolumns{3} 
\tablecaption{GAUSSIAN COMPONENTS IN THE FIT TO THE HI LINE PROFILE \label{table_1}}
\tablewidth{0pc}
\tablehead{
\colhead{S$_l$} & \colhead{V$_{LSR}$}      & \colhead{$\Delta$V} \\
\colhead{(mJy)} & \colhead{(km~s$^{-1}$)}  & \colhead{(km~s$^{-1}$)} }
\startdata
-98$\pm$36& 1.2$\pm$0.6 & 3$\pm$1 \\
-49$\pm$24& 6.4$\pm$1.5 & 5$\pm$3 \\
-69$\pm$27& 11.1$\pm$0.6 & 2$\pm$1 \\
267$\pm$18& 15.5$\pm$0.1 & 4$\pm$1 \\
\enddata


\end{deluxetable}

\end{document}